\documentclass{ws-ijmpd}
\RequirePackage{graphicx}
\usepackage{epstopdf}
\usepackage{bm}
\usepackage{amsfonts}
\usepackage{lineno,hyperref}
\begin{document}

\title{ANISOTROPIC STRANGE STAR WITH TOLMAN $V$ POTENTIAL}

\author{DIBYENDU SHEE}
\address{Department of Physics, Indian Institute of Engineering Science
and Technology, Shibpur, Howrah, West Bengal, 711103, India\\ dibyendu\_shee@yahoo.com}

\author{DEBABRATA DEB}
\address{Department of Physics, Indian Institute of Engineering Science
and Technology, Shibpur, Howrah, West Bengal, 711103, India\\ ddeb.rs2016@physics.iiests.ac.in}

\author{SHOUNAK GHOSH}
\address{Department of Physics, Indian Institute of Engineering Science
and Technology, Shibpur, Howrah, West Bengal, 711103, India\\ shnkghosh122@gmail.com}

\author{SAIBAL RAY}
\address{Department of Physics, Government College of
Engineering \& Ceramic Technology, Kolkata 700 010, West Bengal, India\\ saibal@associates.iucaa.in}

\author{B.K. GUHA}
\address{Department of Physics, Indian Institute of Engineering Science
and Technology, Shibpur, Howrah, West Bengal, 711103, India\\ bkguhaphys@gmail.com}

\maketitle

\begin{history}
\received{Day Month Year} \revised{Day Month Year} \comby{Managing
Editor}
\end{history}

\begin{abstract}
In this paper we present a strange stellar model using Tolman $V$ type metric potential employing simplest form of the MIT bag equation of state (EOS) for the quark matter. We consider that the stellar system is spherically symmetric, compact and made of an anisotropic fluid. Choosing different values of $n$ we obtain exact solutions of the Einstein field equations and finally conclude that for a specific value of the parameter $n=1/2$ we find physically acceptable features of the stellar object. Further we conduct different physical tests, viz., the energy condition, generalized TOV equation, Herrera's cracking concept, etc., to confirm physical validity of the presented model. Matching conditions provide expressions for different constants whereas maximization of the anisotropy parameter provides bag constant. By using the observed data of several compact stars we derive exact values of some of the physical parameters and exhibit their features in a tabular form. It is to note that our predicted value of the bag constant satisfies the report of CERN-SPS and RHIC.
\end{abstract}


\section{Introduction}\label{sec1}
Einstein~\cite{Einstein1916}, in 1916, opened up a new way to look at the universe through his General Theory of Relativity. The striking idea that matter and energy creates curvature in the spacetime was beyond the imagination before him. New branches of physics, known as
Astrophysics and Cosmology, were created on the basis of GTR. Still now this field attracts the researchers to solve the hidden mystery of universe, with the same force as in the initial day. In $1916$ Schwarzschild found out the solution of the Einstein field equations describing a star with uniform matter density. The exact solution of the Einstein field equations are very much relevant to describe the nature of a compact star. Though after that many scientists obtained new exact solutions but very few of them satisfied the physical conditions inside the stellar interior. A survey report, made by Delgaty and Lake~\cite{Delgaty1988}, showed that out of $127$ known solutions only $16$ solutions satisfied all the physical conditions. 

A neutron star is the final stage of a gravitationally collapsing star. It stabilized by degenerate neutron pressure after exhausting all the thermo nuclear fuel inside it. A compact star is more compact than an ordinary neutron star. Some of the examples of such compact stars  are X-ray burster $4U 1820-30$, X-ray pulsar $Her X-1$, X-ray sources RX J $185635-3754$, Milisecond pulsar SAX J $1808.4-3658$, PSR $0943+10$ because their estimated compactness cannot be explained properly in terms of ordinary neutron star Equation of State (EOS)~\cite{Alcock1986,Haensel1986,Weber2005,Perez-Garcia2010,Rodrigues2011,Bordbar2011}. The theoretical modelling of neutron stars have improved considerably over the last few decades after much improvement of the understanding of high energy particle interactions~\cite{Shapiro1983}. Ruderman~\cite{Ruderman1972} had shown that matter densities of compact stars are to be of the order of $10^{15} gm/cc$ or higher. At this high density range all the nuclear interactions have to treat relativistically. So one can expect that anisotropy in pressure may occur and it can be decomposed into two parts namely radial pressure $p_r$ and tangential pressure $p_t$~\cite{Bowers1974,Sokolov1980}. At this high density range anisotropy may occur in various reasons, e.g., existence of a solid core, phase transition, presence of type {\it P}-superfluid, rotation, magnetic field, existence of external field, mixture of two fluids etc.

The renowned particle physists  Gell-Mann~\cite{Gell-Mann1964} and Zweig~\cite{Zweig1964} independently proposed that hadrons are composed even with more fundamental particles known as quarks. Though this quark model get support of experiment later on.
According to Witten~\cite{Witten1984} and Farhi et al.~\cite{Farhi1984} quark matter might be the true ground state of hadron. This idea gives birth of an entirely new class of stellar objects composed of deconfined  {\it u}, {\it d} and {\it s} quarks, known as `quark stars'. In  quantum chromodynamics (QCD) the quark confinement mechanism have been dealt with great details as they are not seen as free particles. According to QCD, for large exchange of momentum the quark interactions becomes weak. At very high temperature
or large density, or at both of the conditions, the quark interactions becomes very weak and as a result they become deconfined. At a high temperature ($\sim $180 Mev or above) deconfinement of quarks are shown in Heavy Ion Collider Experiment. But the temperature of the neutron stars are of the order of few keV. So for deconfinement of the quarks at the centre of the neutron star, an extremely large chemical potential is required. The quark star results if the neutron star matter get converted totally into quark matter. Theoretically it is possible, under some conditions, that some of the up and down quarks transformed into strange quarks. Since strange matter is the true ground state of matter so quark star gets the thrust to converted totally into a strange star if its core once converted into a strange matter. Therefore a neutron star gets converted into strange star.

The strange matter can be formed by the following two ways, as proposed by Witten~\cite{Witten1984} that (i) at ultrahigh density range neutron star converted into a strange star and (ii) as a result of quark-hadron phase transition in the early universe.
Actually the transition from the normal hadronic matter to strange matter occurs at very high densities or corresponding to low
temperatures. Cheng et al.~\cite{Cheng1998} proposed that, nearly equal number of up, down and strange quarks are present in the Beta-Equilibrated strange quark star matter. Though there is a slight deficit of the latter. Bodmer~\cite{Bodmer1971}
suggested that, beyond nuclear density a phase transition between hadronic and strange quark matter could occur
in the universe when a massive star explodes as a supernova. As a result the inner core of the neutron star consists
most likely strange quark matter. In general if the mass to size ratio of a superdense star exceeds
$0.3$ then it can be expected to be composed of strange matter~\cite{Tikekar2007}. According to Alford~\cite{Alford2001},
in the dense core of a neutron star there is sufficiently high density and corresponding low temperature to crush the hadrons into quark matter. According to MIT bag model the quark confinement is due to the universal pressure $B_g$, called the Bag Constant. This simple model describes successfully the observations in particle physics. It is actually the difference between energy density of the perturbative and non perturbative quantum chromodynamics vacuum. Farhi and Jaffe~\cite{Farhi1984} as well as Alcock et al.~\cite{Alcock1986} proposed that the value of the bag constant should be within the following range $55-75$ $Mev/{fm}^{3}$, for a stable strange quark matter. In the nucleonic EOSs BSk $19$, BSk $20$ and BSk $21$ Chamel et al.~\cite{Chamel2013} used the values of effective bag Constant to be $78.6$, $65.5$ and $56.7$ $Mev/{fm}^{3}$. The data set of CERN-SPS and RHIC~\cite{Burgio2002} shows that a wide range of bag constant are permissible. 

Tolman in $1939$~\cite{Tolman1939} gave explicit analytic solution of the Einstein field equations by choosing eight different types of metric potentials. Literature survey shows that there are no research works available in the field of astrophysics with the Tolman $V$ metric potential. As anisotropy is the most general case to investigate strange star so such situation has motivated us to study anisotropic strange star with Tolman $V$ metric potential through MIT bag model.

The planning of this paper is as follows: in Sec.~\ref{sec2} we have written down the basic field equations with the help of MIT bag model and Tolman $V$ metric potential. In Sec.~\ref{sec3} we have obtained solutions to the Einstein field equations along with the expressions for the anisotropy and EOS parameter whereas in Sec.~\ref{sec4} from the boundary condition we have found out the values of the constants. Sec.~\ref{sec5} can be decomposed into many subsections~\ref{subsec5.1} to~\ref{subsec5.4} containing the energy conditions, generalized TOV equation, Herrera's cracking concept and adiabatic index respectively. The mass-radius relation of the star and redshift are considered in Sec.~\ref{sec6}. Finally some concluding remarks are provided in Sec.~\ref{sec7}.

\section{Basic field equations of Einstein's space-time}\label{sec2}
The static spherically symmetric spacetime can be described by the line element
\begin{equation}
{ds}^{2}=-{{e}^{\nu(r)}}{{dt}^{2}}+{{e}^{\lambda(r)}}{{dr}^{2}}+{r}^{2}({{d\theta}^{2}}+{{sin}^{2}}\theta{{d\phi}^{2}}),\label{2.1}
\end{equation}
where $\nu(r)$ and $\lambda(r)$ are the metric potentials and function of the radial coordinate only. These metric coefficients have much
significance which actually help us to realize the gravitational environment of the stellar system. Considering the requirement for our study 
we have assumed the form of line element which is most general and spherically symmetric. Here basically $g_{00} = e^{\nu}$ and $g_{11} = e^{-\lambda}$ such that only the magnitudes of these metric coefficients can change completely the scenario of physical structure~\cite{Maurya2017}.

The most general energy-momentum tensor for anisotropic matter distribution is given by
\begin{eqnarray}
{T_{{\nu}}}^{\mu}= (+\rho, - p_r, - p_t, - p_t), \label{2.2}
\end{eqnarray}
where $\rho$ is the matter density, $p_r$ is the radial pressure and $p_t$ is the
tangential pressure of the fluid which is in the orthogonal direction to $p_r$. 

One can obtain the Einstein field equations, assuming $G=1=c$, as follows
\begin{eqnarray}\label{2.3}
& \qquad{{\rm e}^{-\lambda}} \left( {\frac {{\lambda}^{\prime}}{r}}-\frac{1}{r^{2}} \right) +{\frac{1}{{r}^{2}}}=8\pi 
\rho,\\ \label{2.4}
& \qquad {{\rm e}^{-\lambda}}\left({\frac{{\nu}^{\prime}}{r}+\frac {1}{r^{2}}}\right)-{\frac{1}{{r}^{2}}}=8\pi{p_r},\\ \label{2.4a}
& \qquad\hspace{-0.5cm} \frac{1}{2}{{\rm e}^{-\lambda}}\left[\frac{1}{2}{\left({\nu}^{\prime}\right)}^{2}+{\nu}^{\prime\prime}-
\frac{1}{2}{\lambda}^{\prime}{\nu}^{\prime}+\frac{1}{r}\left({\nu}^{\prime}-{\lambda}^{\prime}\right)\right]=8\pi{p_t},
\end{eqnarray}
where prime denotes differentiation with respect to the radial coordinate $r$. 

As we have mentioned in the Introduction that we are dealing with strange star in this paper so the equation of state, according to MIT bag model, will be 
\begin{equation}
p_r=\frac{1}{3}\left(\rho-4{B_g}\right),\label{2.5}
\end{equation}
where $B_g$ is bag constant.

We define the mass function $m(r)$ of the star as
\begin{equation}
m \left( r \right) =4\pi \int_{0}^{r}\rho \left( r \right) {r}^
{2}{dr}.\label{2.6}
\end{equation}

Now to solve the field equation we choose Tolman V~\cite{Tolman1939} potential 
\begin{equation}
{{\rm e}^{\nu}}=K{r}^{2n}, \label{2.7}
\end{equation}
where $K$ is the constant and $n$ is a parameter whose numerical value as chosen by Tolman~\cite{Tolman1939} was $1/2$. The choice of the above Tolman $V$ type potential is not all arbitrary here. Actually we wanted to study the case of strange star for different values of $n$, so that we could fix up the range of the index parameter $n$. However, after performing several numerical values of $n$ we have found that the model is stable only for $n=1/2$. 

Another metric potential, needed for the spherical symmetry, is given by
\begin{equation}
{{\rm e}^{-\lambda}}=1-{\frac {2m}{r}}, \label{2.8}
\end{equation}
and this will help us to calculate the compactification factor in further.

\section{Solution of the Einstein field equations}\label{sec3}
Now from Eqs.~(\ref{2.3}),~(\ref{2.4}),~(\ref{2.5}) and~(\ref{2.7}) we find
\begin{eqnarray}
& \qquad\hspace{-1cm} {{\rm e}^{\lambda}}=-{\frac {3{r}^{4} \left( 3{n}^{2}+5n+2
 \right) {r}^{6n}}{ \left[\left( 48{r}^{6}\pi B_{{g}}-6{r}^{
4} \right) n+32{r}^{6}\pi B_{{g}}-6{r}^{4} \right] {r}^{6n}+9c_{{1}} \left( n+1 \right)\left( n+\frac{2}{3}\right) }}, \nonumber \\ \label{2.9}
\end{eqnarray}
where $c_{1}$ is an integration constant whose value can be determined from the boundary condition. For any physically viable model the metric potentials should be finite and free from singulrities inside the stellar system. Our model satisfies these conditions as at the centre we get  ${e^{\nu}|_{r=0}}=0$ and ${e^{\lambda}}|_{r=0}>1$. 

Using the Einstein field equations~(\ref{2.3})-(\ref{2.4a}) and Eqs.~(\ref{2.5})-(\ref{2.7}) and (\ref{2.9}) we find the following physical parameters which are given by
\begin{eqnarray}\label{4.1}
& \qquad\hspace{-0.5cm} \rho={\frac {\rho_{{1}}{r}^{-\left(4+6n\right)}+3{n}^{2}+ \left( 48B_{{g
}}\pi {r}^{2}+3 \right) n+32B_{{g}}\pi {r}^{2}}{ 8\left( 3{n}^{
2}+5n+2 \right) \pi {r}^{2}}},\\ \label{4.2}
& \qquad\hspace{-0.5cm} p_{{r}}={\frac {\rho_{{1}}{r}^{-\left(4+6n\right)} + \left(3 - 96B_{{g}}\pi
{r}^{2} \right) {n}^{2}+ \left(3 -112B_{{g}}\pi {r}^{2}
 \right) n-32B_{{g}}\pi {r}^{2}}{ 24\left( 3{n}^{2}+5n+2
 \right) \pi {r}^{2}}},  \label{4.3} \nonumber \\ \\
& \qquad\hspace{-0.5cm} p_{{t}}={\frac {-\rho_{{1}} \left( n+2 \right) {r}^{-4-6n}+p_{
{1}}-16B_{{g}}\pi {r}^{2} \left( 5n+2 \right) }{24 \left( 3{n}^{
2}+5n+2 \right) \pi {r}^{2}}}, \label{4.4}
\end{eqnarray}
where\\
$\rho_{{1}}=-36 \left( n+\frac{1}{2}\right)  \left[ \rho_{{2}}- \left( \frac{1}{2}
{n}^{2}+\frac{n}{2}\right) {R}^{6n+4}-\rho_{{3}} \right] $, \\ 
$\rho_{{2}}= M \left( n+1 \right) \left( n+\frac{2}{3}\right) {R}^{6n+3}$, \\
$\rho_{{3}}=\frac{8}{3}\left( n+\frac{2}{3} \right) B_{{g}}\pi {R}^{6n+6}$, \\
$p_{{1}}= \left( 6-48B_{{g}}\pi {r}^{2} \right) {n}^{3}+ \left( 6-80B_{{g}}\pi {r}^{2} \right) {n}^{2}$.

\begin{figure}[!htp]
\centering
    \includegraphics[width=6cm]{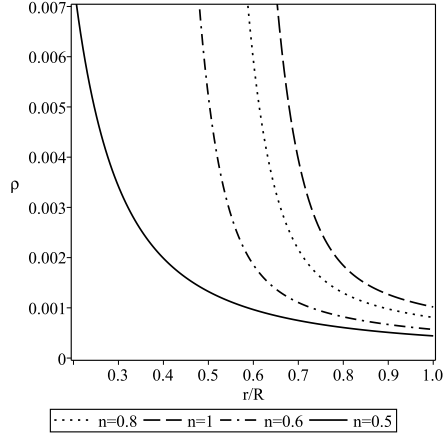}
        \caption{Variation of the density with the fractional radial coordinate $r/R$ for $LMC\,X-4$} \label{rho}
\end{figure}

\begin{figure}[!htp]
\centering
    \includegraphics[width=6cm]{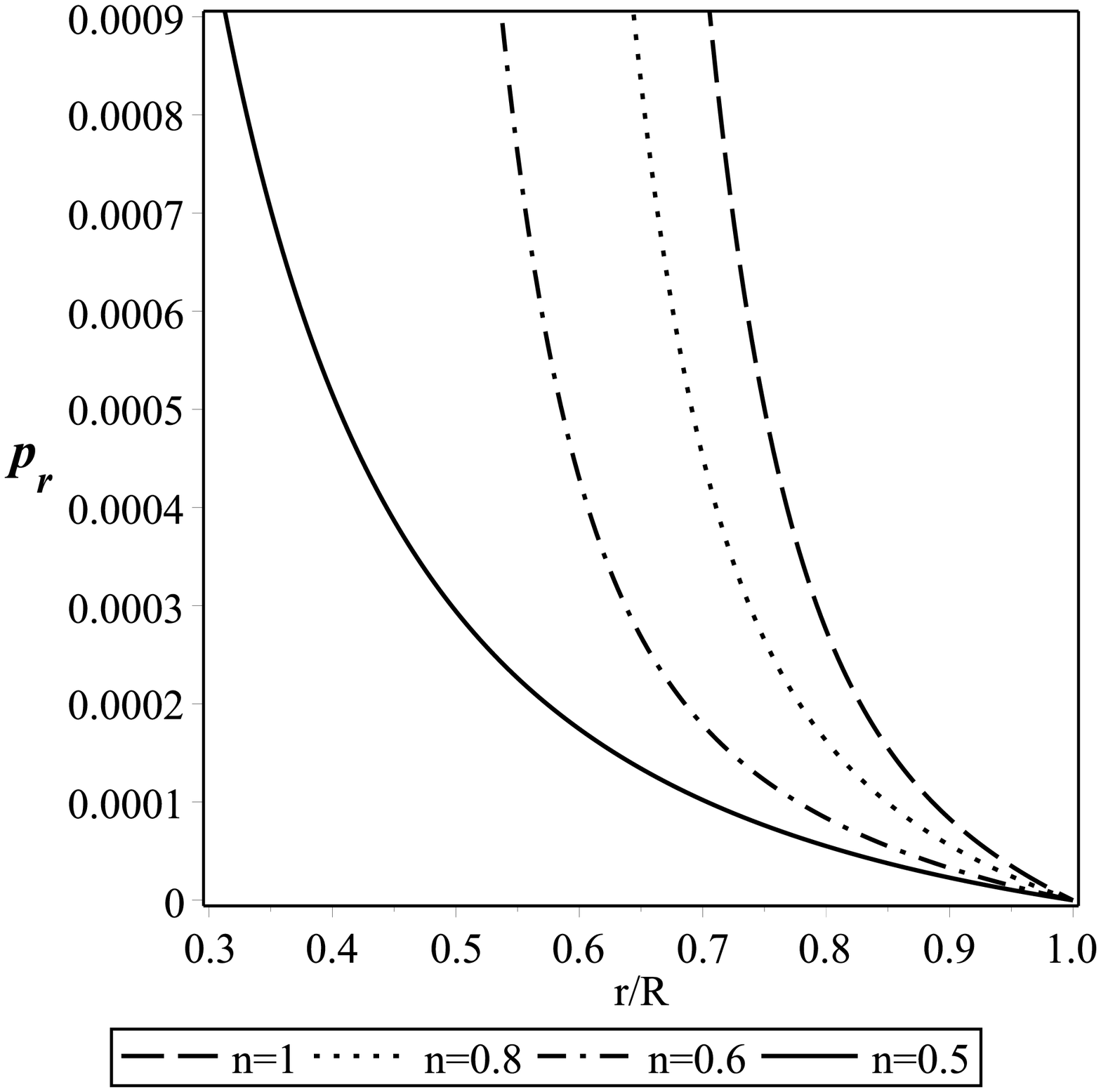}
    \includegraphics[width=6cm]{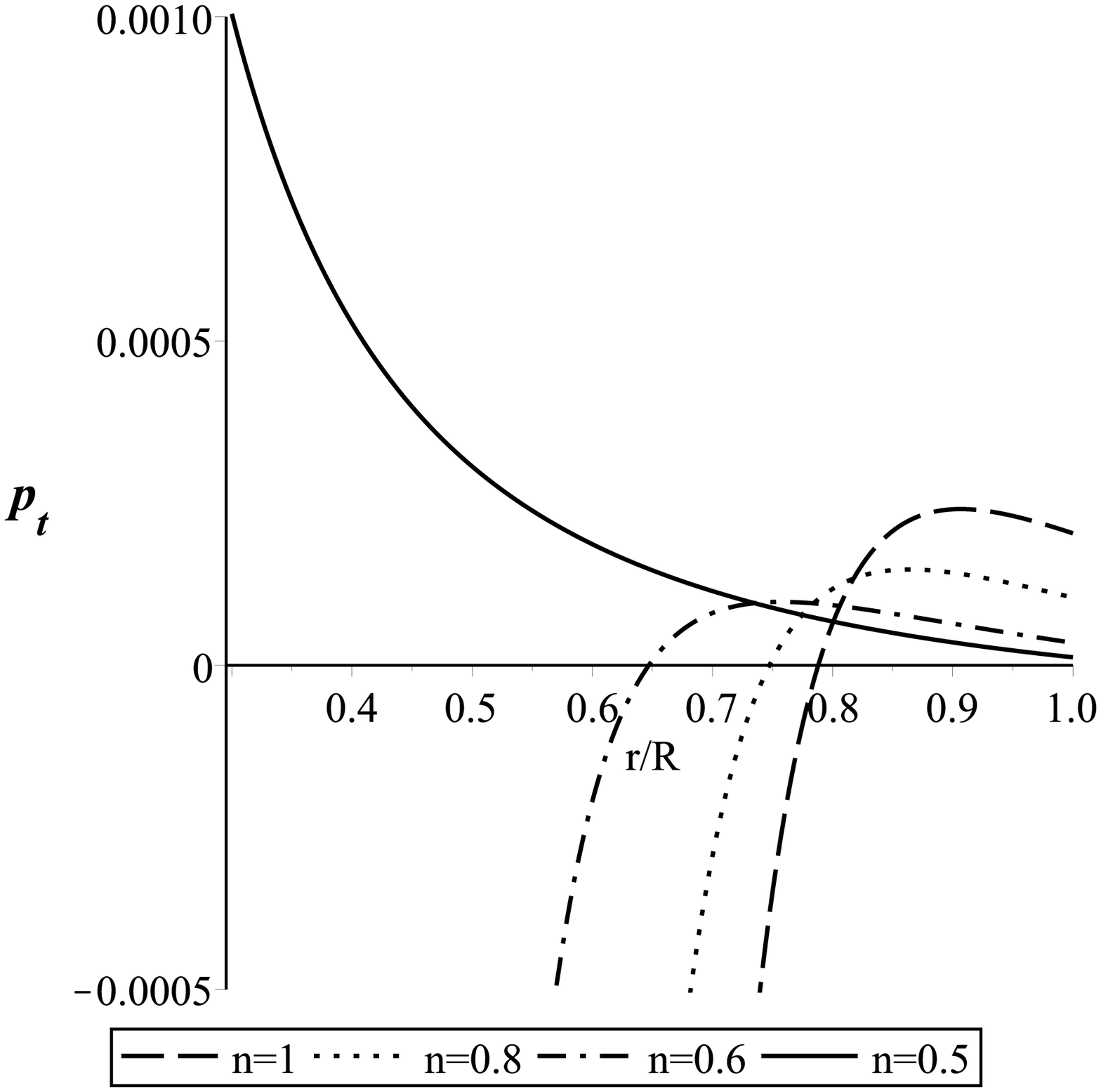}
        \caption{Variation of the radial $p_r$ (left panel) and $p_t$ (right panel) with the fractional radial coordinate $r/R$ for $LMC\,X-4$} \label{pres}
\end{figure}

The variations of density and pressures are shown in Fig.~\ref{rho} and~\ref{pres}. From both the Figs.~\ref{rho} we see that the matter density, radial pressure and tangential pressure all decrease monotonically. In the case of density we note from Fig~\ref{rho} that $\rho(0)  \rightarrow \infty$ and $\rho(R) \rightarrow $ finite. Also from Fig.~\ref{pres} it is clear that for only $n=1/2$ we are getting positive value of the tangential pressure inside the stellar configuration and henceforth we shall consider only the $n=1/2$ case for further physical study.

The anisotropy of the system is calculated as follows
\begin{eqnarray}
& \qquad\hspace{-1cm} \Delta={\frac {-\rho_{{1}} \left( n+3 \right) {r}^{-(4+6n)}+
 \left( 96B_{{g}}\pi {r}^{2}-3 \right) {n}^{2}+ \left( 32B_{{g}}
\pi {r}^{2}-3 \right) n+p_{{1}}}{ 24\left( 3{n}^{2}+5n+2 \right)
\pi {r}^{2}}}. \nonumber \\\label{4.5}
\end{eqnarray}

\begin{figure}[!htp]
\centering
    \includegraphics[width=6cm]{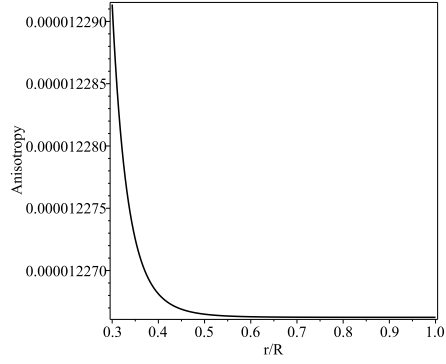}
            \caption{Variation of the anisotropy with the fractional radial coordinate $r/R$ for $LMC\,X-4$} \label{aniso}
\end{figure}

The term $2\Delta/r$ is called anisotropic force, which will be directed outward (i.e. repulsive in nature) if $\Delta>0$
and directed inward (i.e. attractive in nature) if $\Delta<0$. The Fig.~\ref{aniso} tells us that $\Delta>0$ throughout the stellar distribution. So here $p_{t}>p_{r}$, i.e., $\Delta>0$ will helps to construct the more compact object according to Gokhroo and Mehra~\cite{Gokhroo1994}. 

Now for a physically valid stellar model the radial pressure must vanish at the surface $(r=R)$, i.e., ${p_r}\left(R\right)=0$ and we have
\begin{eqnarray}
n={\frac {M}{R-2M}}.\label{4.6}
\end{eqnarray}

In the above Eq.~(\ref{4.6}) for finite value of the parameter $n$ we impose the condition $R \neq 2M$. However, it is interesting to note that at the radius $R$, if $R = 2M$, one can get $n \rightarrow \infty$. This therefore provides a limiting value of $n$ to avoid any unphysical situation. In the present study we have opted for the value of $n=1/2$ as other values do not yield physically viable results. This immediately suggests the radius of the present compact star as $R = 4M$ and can be written in the form $2M/R=0.5$. This fulfils the Buchdahl condition $2M/R<0.88$~\cite{Buchdahl1959} in connection to the mass-radius relationship which will be discussed later on in details.  

Following Deb et al.~\cite{Deb2016} we maximize anisotropy at the surface and find the bag constant as
\begin{eqnarray}\label{4.7}
& \qquad\hspace{-0.5cm} B_{{g}}={\frac {12M{n}^{3}-6{n}^{3}R+54M{n}^{2}-23{n}^{2
}R+60Mn-14nR+18M}{16 \left( 2{n}^{2}+7n+3 \right) \pi {R}^{3
}}}.\nonumber \\
\end{eqnarray}

The radial~$\left({{\omega}_r}\right)$ and tangential~$\left({\omega}_t\right)$  EOS parameters for our system can be written in the following form as
\begin{eqnarray}\label{3.7}
& \qquad\hspace{-7.2cm} \omega_{{r}}=\frac{p_r}{\rho}\nonumber \\
& \qquad\hspace{-0.6cm}={\frac {\rho_{{1}}{r}^{-\left(4+6n\right)}+ \left(3-96B_{{g}}
\pi {r}^{2} \right) {n}^{2}+ \left(3-112B_{{g}}\pi {r}^{2}
 \right) n-32B_{{g}}\pi {r}^{2}}{3\rho_{{1}}{r}^{-\left(4+6n\right)}+3{n}^{2
}+ \left(48B_{{g}}\pi {r}^{2}+3 \right) n+32B_{{g}}\pi {r}^{2}}}, \nonumber \\ 
\end{eqnarray}

\begin{figure}[!htp]
\centering
    \includegraphics[width=6cm]{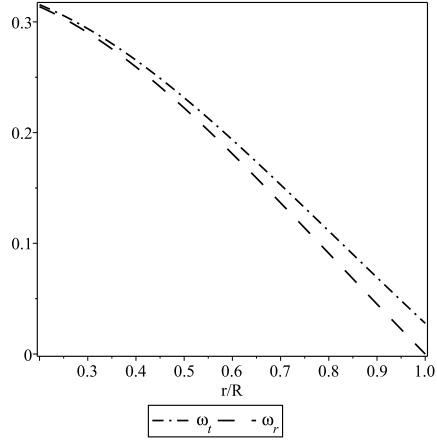}
           \caption{Variation of the radial and transverse EOS parameter with the fractional radial coordinate $r/R$ for $LMC\,X-4$} \label{eos}
\end{figure}

\begin{eqnarray}\label{3.8}
& \qquad\hspace{-7.2cm} \omega_{{t}}=\frac{p_t}{\rho}\nonumber \\
& \qquad\hspace{-0.6cm}={\frac {-\rho_{{1}} \left( n+2 \right) {r}^{-\left(4+6n\right)}-80
B_{{g}}\pi n{r}^{2}-32B_{{g}}\pi {r}^{2}+p_{{1}}}{3\rho_{{1}}{
r}^{-\left(4+6n\right)}+9{n}^{2}+ \left( 144B_{{g}}\pi {r}^{2}+9 \right) n+
96B_{{g}}\pi {r}^{2}}}.
\end{eqnarray}

From Fig.~\ref{eos}, we get $0 \leq (\omega_r, \omega_t) \leq 1/3$ where maximum value of EOS starts from the center and decreases to zero at the surface of the spherical stellar system. Moreover, we note that here $\omega_r < \omega_t$.

\section{Matching boundary conditions}\label{sec4}
Now we shall match our interior solutions to the exterior Schwarzschild metric at the boundary $r=R$, here $R$ being the total radius of the star, given as
\begin{eqnarray}    \label{3.1}
&\qquad\hspace{-1cm} ds^{2} =-\left(1-\frac{2M}{r} \right)\, dt^{2} + \frac{dr^{2}}{\left(1-\frac{2M}{r}\right)}+r^{2} (d\theta ^{2}+\sin ^{2} \theta \, d\phi ^{2} ),\nonumber \\
\end{eqnarray}

\begin{figure}[!htp]
\centering
    \includegraphics[width=6cm]{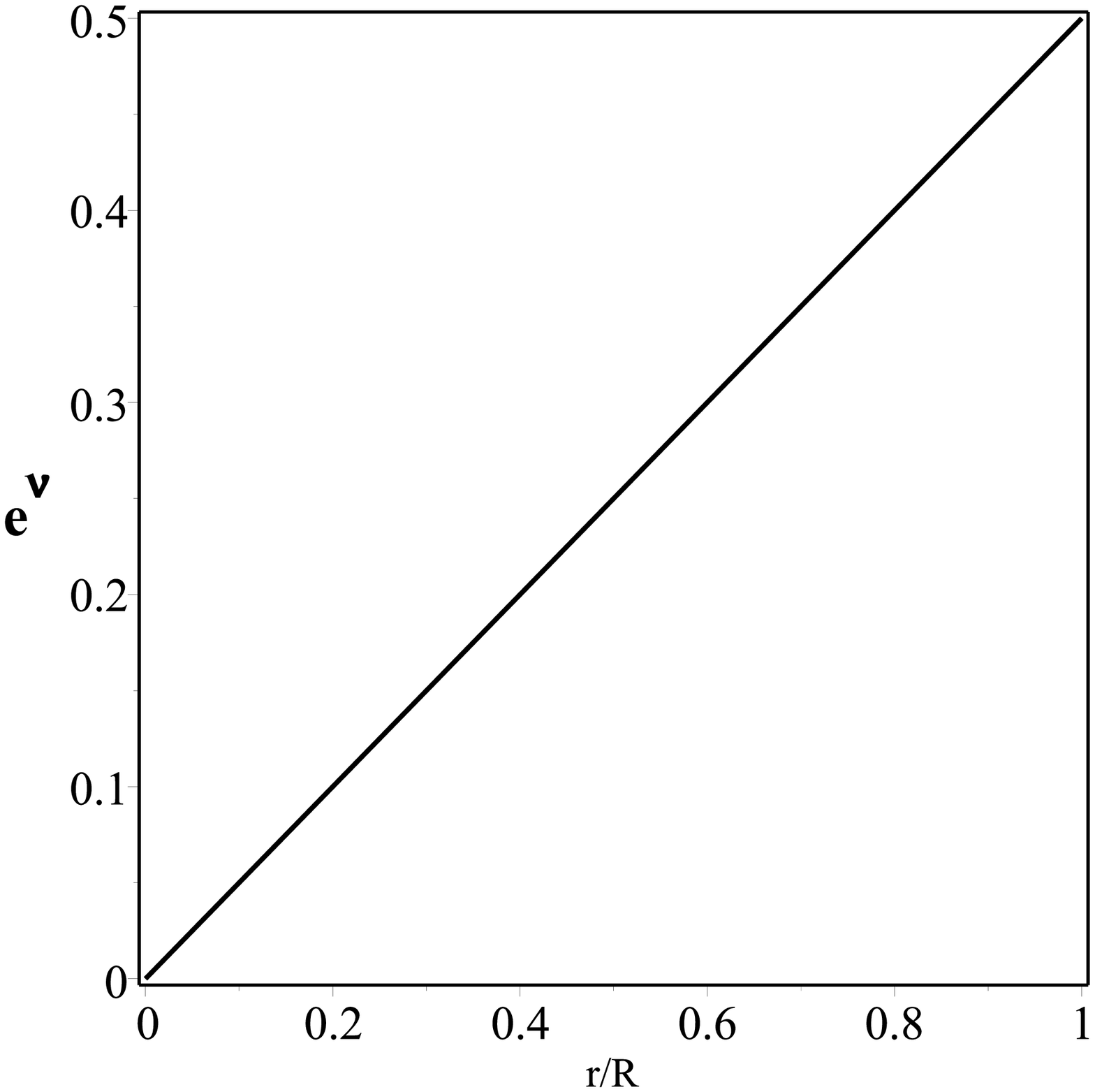}
    \includegraphics[width=6cm]{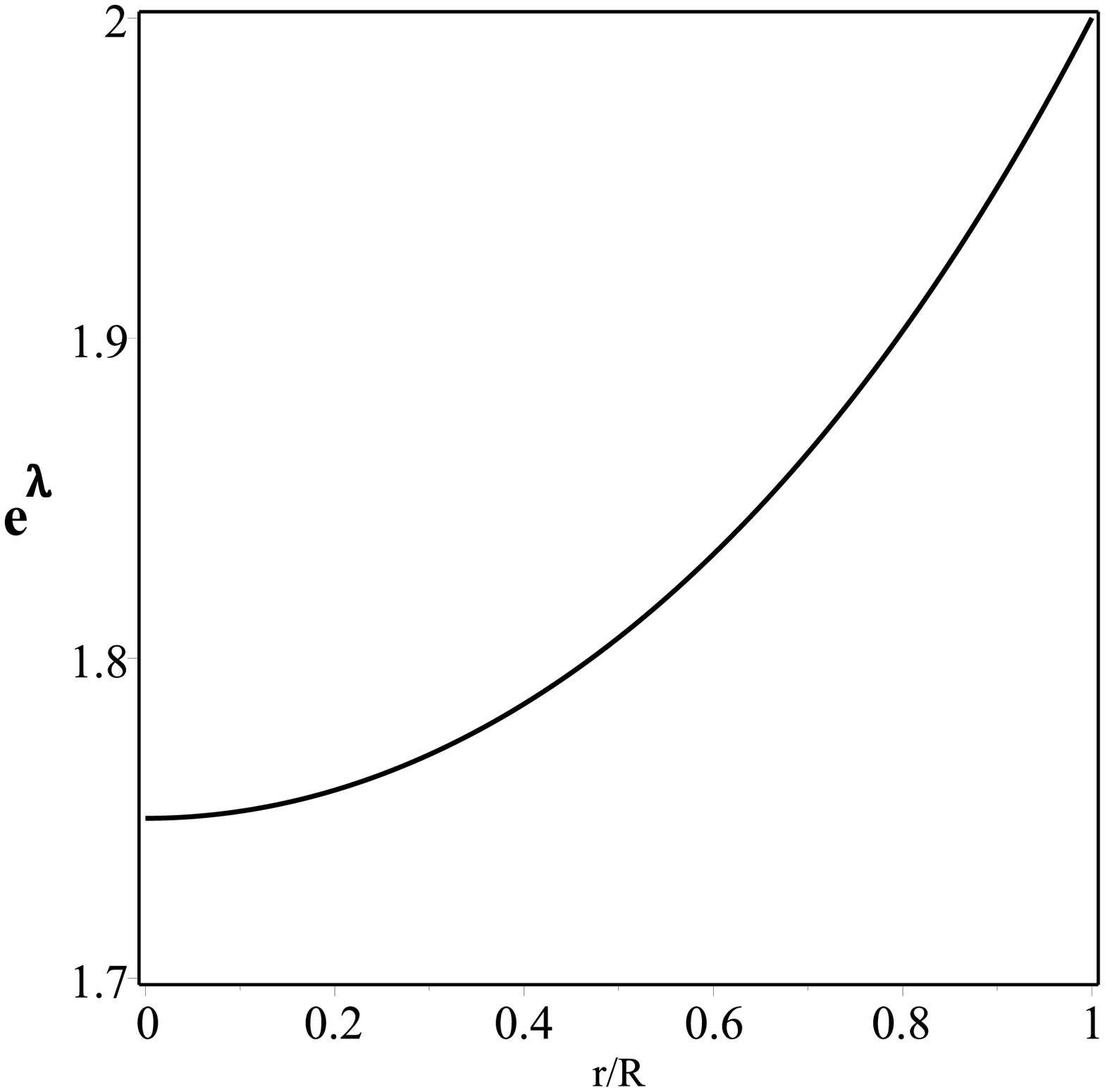}
        \caption{Variation of ${\rm e}^{\nu}$ (left panel) and ${\rm e}^{\lambda}$ (right panel) with the fractional radial coordinate $r/R$ for $LMC\,X-4$ for $n=1/2$} \label{pot}
\end{figure}

At the boundary the coefficients $g_{rr}$, $g_{tt}$ and $\frac{\partial g_{tt}}{\partial r}$ all are continuous and the continuity of $g_{rr}$, $g_{tt}$ gives us (vide Fig.~\ref{pot})
\begin{eqnarray}\label{3.2}
& \qquad\hspace{-4cm} {{\rm e}^{\nu \left( R \right) }}=K{R}^{2n}=1-{\frac {2M}{R}},\\ \label{3.3}
& \qquad {{\rm e}^{\lambda \left( R \right) }}=-{\frac {3{R}^{4} \left( 3{n
}^{2}+5n+2 \right) {R}^{6n}}{9c_{{1}} \left( n+1 \right)\left( n+\frac{2}{3} \right)+48{R}^{4} \left[ \pi B_{{g}}
 \left( n+\frac{2}{3} \right) {R}^{2}-\frac{n+1}{8}\right] {R}^{6n}}} \nonumber \\
& \qquad\hspace{-4cm} ={\left(1-{\frac {2M}{R}}\right)}^{-1}.
\end{eqnarray}
From the above two equations we will get the value of the constant $K$ and integration constant $c_{1}$ which are given by
\begin{eqnarray}\label{3.4}
& \qquad\hspace{-2cm} K= \frac{1}{R^{2n}}{\left(1-{\frac {2\,M}{R}}\right)},\\ \label{3.5}
& \qquad\hspace{-0.5cm} c_{{1}}={\frac {-48\, \left( n+\frac{2}{3}\right) B_{{g}}\pi {R}^{6n+6}-9
 \left( n+1 \right)  \left[ -2 \left( n+\frac{2}{3} \right) M{R}^{6n+3}+
{R}^{6n+4}n \right] }{9{n}^{2}+15n+6}}. \nonumber \\
\end{eqnarray}

We shall use numerical values of these constants to plot the graphs.

\section{Physical properties of the stellar model}\label{sec5}

\subsection{Energy conditions}\label{subsec5.1}
General relativity permits energy-momentum tensor $T^{ij}$ to describe the distribution of mass, momentum and stress due to matter and to any non-gravitational fields. Though Einstein's field equation not directly concern with what kind of states of matter or non-gravitational fields are admissible in the spacetime, the energy conditions allows all states of matter and all non-gravitational fields in GTR and ruled out many unphysical solutions. Therefore GTR allows the following energy conditions known as the Null Energy Condition (NEC), Weak Energy Condition (WEC), Strong Energy Condition (SEC) and 
Dominant Energy condition (DEC). 

 The energy conditions are satisfied if and only if the following inequalities hold simultaneously by every point inside the fluid sphere:
 \begin{eqnarray}\label{5.1.1}
& \qquad NEC: \rho \geq 0,\\ \label{5.1.2}
& \qquad {WEC_r}:\rho+{p_r} \geq 0, \rho\geq 0, \\ \label{5.1.3}
& \qquad {WEC_t}:\rho+{p_t} \geq 0, \rho\geq 0, \\ \label{5.1.4}
& \qquad {SEC}:  \rho+{p_r} \geq 0, \rho+{p_r}+2{p_t}> 0,\\ \label{5.1.5}
& \qquad {DEC_r}: \rho>|p_r|,~{DEC_t}: \rho>|p_t|. \label{5.1.6}
\end{eqnarray}

\begin{figure}[!htp]
\centering
    \includegraphics[width=7cm]{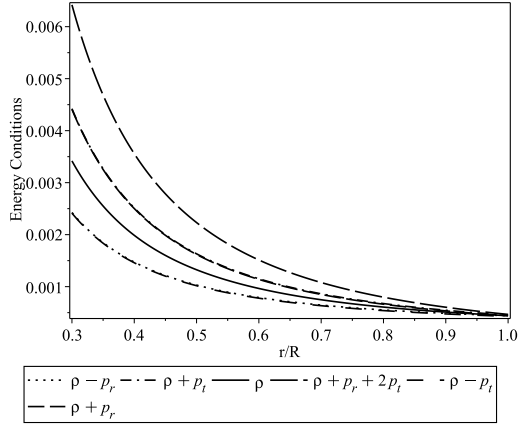}
           \caption{Variation of the energy conditions with the fractional radial coordinate $r/R$ for $LMC\,X-4$} \label{energy}
\end{figure}

From Fig.~\ref{energy} it is observed very clearly that NEC, WEC, SEC and DEC are satisfied by our model.

\subsection{Generalized TOV equation}\label{subsec5.2}
To search equilibrium situation of this anisotropic star under different forces, the generalized Tolman-Oppeheimer-Volkoff (TOV) equation
can be written as~\cite{Varela2010}
\begin{equation}
-\frac{M_G(\rho+p_r)}{r^{2}}e^{\frac{\lambda-\nu}{2}}-\frac{dp_r}{dr}+\frac{2}{r}(p_t-p_r)=0,\label{5.2.1}
\end{equation}
where $M_G=M_G(r)$ is the effective gravitational mass inside a sphere of radius $r$ which can be derived from the modified 
Tolman-Whittaker formula~\cite{Devitt1989} as
\begin{equation}
M_G(r)=\frac{1}{2}r^{2}e^{\frac{\nu-\lambda}{2}}\nu'.\label{5.2.2}
\end{equation}
substituting this value in Eq.~(\ref{5.2.1}) we get the following form of TOV equation
\begin{equation}
-\frac{\nu'(\rho+p_r)}{2}-\frac{dp_r}{dr}+\frac{2}{r}(p_t-p_r)=0,\label{5.2.3}
\end{equation}
which can be written in the following form as
\begin{equation}
F_g+F_h+F_a=0,\label{5.2.4}
\end{equation}
where $F_g$, $F_h$ and $F_a$ represents respectively gravitational force, hydrostatic force and anisotropic force.

From our model we get the following expressions of the above mentioned terms
\begin{eqnarray}\label{5.2.5}
&\qquad\hspace{-7cm} {F_g}=-\frac{\nu'}{2}(\rho+p_r)\nonumber \\
&\qquad\hspace{-1cm}=-{\frac { \left( -24B_{{g}}\pi {n}^{2}{r}^{2}+8B_{{g}}\pi
n{r}^{2}+16B_{{g}}\pi {r}^{2}+\rho_{{1}}{r}^{-(4+6n)}+3{n}^{2}
+3n \right) n}{6{r}^{3} \left( 3{n}^{2}+5n+2 \right) \pi }}\nonumber \\ \\\label{5.2.6}
&\qquad\hspace{-5.6cm} {F_h}=-\frac{dp_r}{dr}={\frac {n+\rho_{{1}}{r}^{-\left(4+6n\right)}}{4\pi \left( 3n+2 \right) {
r}^{3}}}\\\label{5.2.7}
&\qquad\hspace{-7.2cm} {F_a}=\frac{2}{r}(p_t-p_r)\nonumber \\
&\qquad\hspace{-0.9cm}={\frac {-\rho_{{1}} \left( n+3 \right) {r}^{-(4+6n)}+ \left( 96
B_{{g}}\pi {r}^{2}-3 \right) {n}^{2}+ \left( 32B_{{g}}\pi {r}^
{2}-3 \right) n+p_{{1}}}{12{r}^{3} \left( 3{n}^{2}+5n+2 \right) \pi}} \nonumber \\
\end{eqnarray}

\begin{figure}[!htp]
\centering
    \includegraphics[width=6cm]{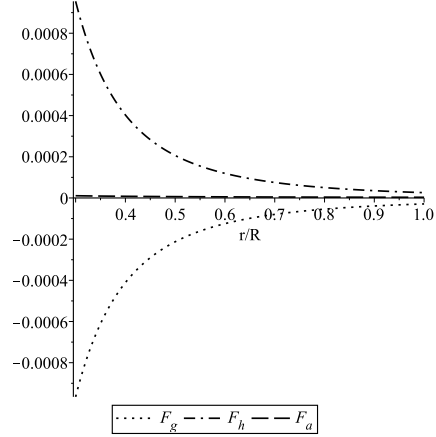}
           \caption{Variation of different forces with the fractional radial coordinate $r/R$ for $LMC\,X-4$} \label{forces}
\end{figure}

From Fig.~\ref{forces} it is obvious that the combined effect of the hydrostatic force and anisotropic force is balanced by the gravitational
force. This figure indicates that our stellar model is in equilibrium under these forces. However, here the anisotropic force is very small with respect to the other two forces. This implies that the radial and tangential pressures do differ in very small amount. Here to achieve stability for the system the hydrodynamic force is taking a major role to counter balance the gravitational force. But still there is a role of anisotropic force which is also balancing the gravitational force. 

\subsection{Herrera cracking concept}\label{subsec5.3}
For a physically realistic model the speed of sound should follow the condition $0\leq{v_{rs}^{2}}\leq1$ and $0\leq{v_{ts}^{2}}\leq1$, which can be termed as causality condition. Also for the stability checking there is an another technique for anisotropic matter distribution, which is known as cracking concept of Herrera~\cite{Herrera1992}. According to this concept the region for which transverse speed of sound is smaller than the radial speed of sound is a potentially stable region~\cite{Herrera1979,Herrera1992,Chan1993,Abreu2007}. The square of radial~$\left({v}_{\it rs}^2\right)$ and tangential~$\left({v}_{\it ts}^2\right)$ sound speed for our system are given as
\begin{eqnarray}\label{5.3.1}
&\qquad\hspace{-6.5cm} {{v}_{\it rs}^2}=\frac{dp_r}{d\rho}=\frac{1}{3},\\ \label{5.3.2}
&\qquad\hspace{-0.5cm} {{v}_{\it ts}^2}=\frac{dp_t}{d\rho}={\frac {-v_{{1}} \left( n+2 \right) +v_{{3}}{R}^{
6n+4}+v_{{2}} \left( n+2 \right) -2{n}^{2}{r}^{6n+4}}{3v_{{1}}
+ \left( -54{n}^{3}-81{n}^{2}-27n \right) {R}^{6n+4}-3v_{{2}
}-3{r}^{6n+4}n}}, \nonumber \\
\end{eqnarray}
where \\
$v_{{1}}=36M \left( n+1 \right)  \left( n+ \frac{1}{2}\right)  \left( n+ \frac{2}{3}\right) {R}^{6n+3}$, \\
$v_{{2}}=96B_{{g}}\pi \left( n+\frac{1}{2} \right)  \left( n+\frac{2}{3} \right) {R}^{6n+6}$, \\
$v_{{3}}=18{n}^{4}+63{n}^{3}+63{n}^{2}+18n$.

\begin{figure}[!htp]
\centering
    \includegraphics[width=6cm]{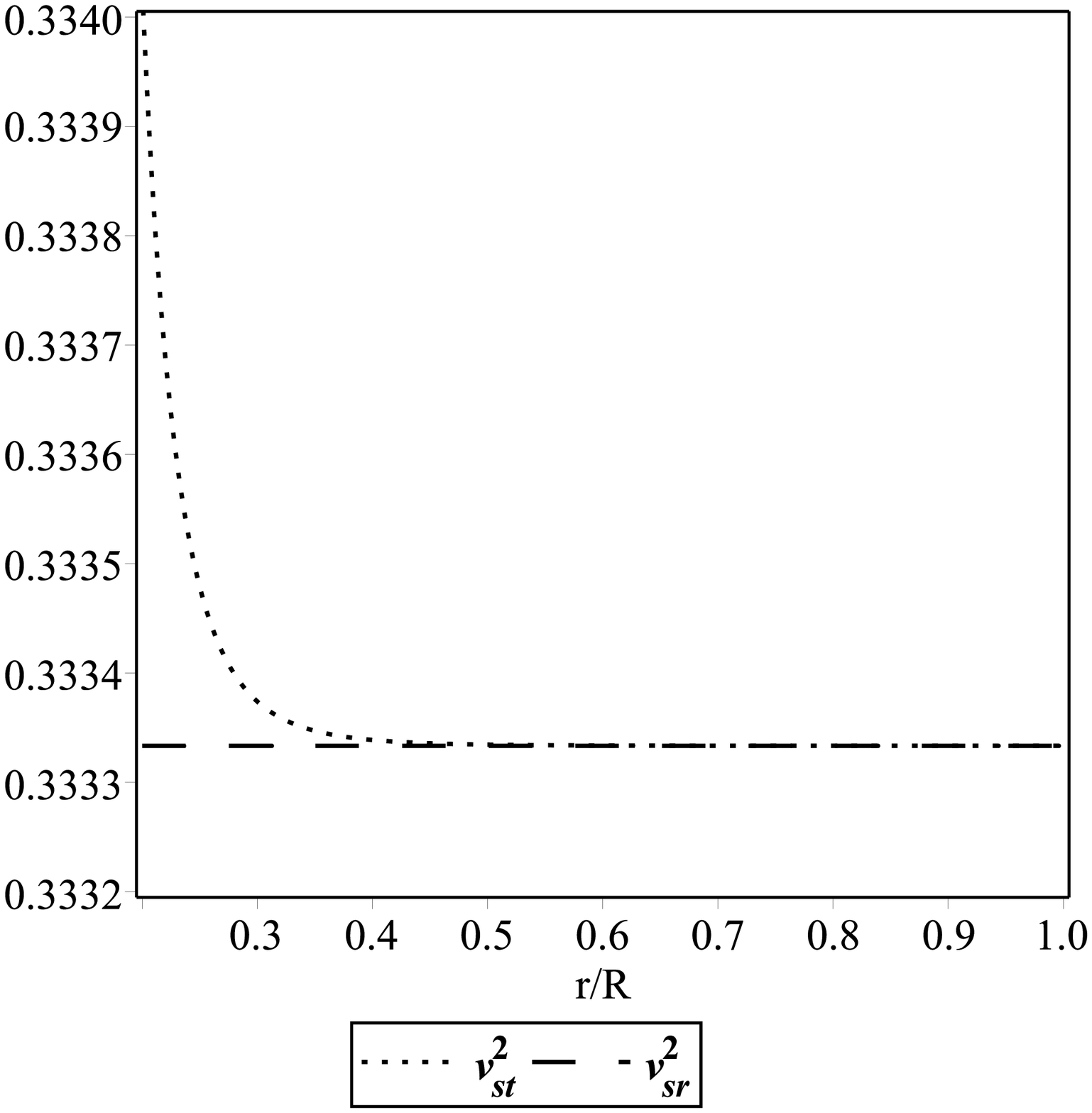}
    \includegraphics[width=6cm]{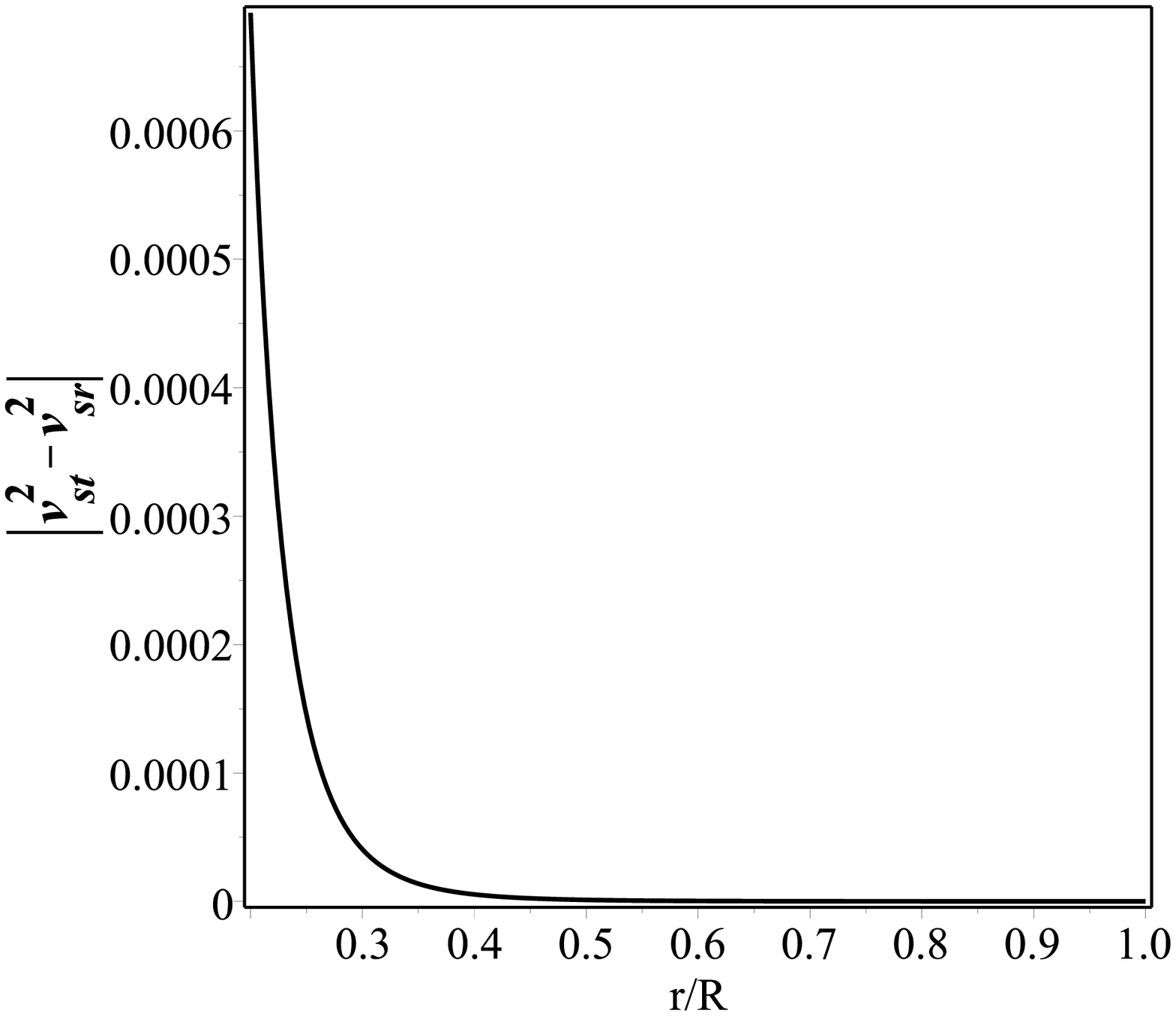}
           \caption{Variation of radial speed of sound, transverse speed of sound (left panel) and velocity difference (right panel) with the fractional radial coordinate $r/R$ for $LMC\,X-4$} \label{vel}
\end{figure}

From Fig.~\ref{vel} it is very clear that through out the stellar structure causality condition and Herrera's cracking condition holds simultaneously. So our model gives a stable configuration.

\subsection{Adiabatic Index}\label{subsec5.4}
For a Newtonian isotropic fluid sphere the condition for stability is given by $\Gamma>\frac{4}{3}$. For a relativistic isotropic sphere it changes and for an anisotropic general relativistic sphere $\Gamma$, known as adiabatic index, can be written as $\Gamma_r$ and
$\Gamma_t$~\cite{Bondi1964}.

For a relativistic anisotropic fluid sphere the stability condition is given by
\begin{eqnarray}
&\qquad \Gamma_r>\frac{4}{3}, \\
&\qquad \Gamma_t > \frac{4}{3}+\left[\frac{4(p_{t0}-p_{r0})}{3|p_{r0}^{'}r|}+\frac{8\pi r \rho_0 p_{r0}}{3|p_{r0}^{'}|}\right],\label{5.4.1}
\end{eqnarray}
where $p_{r0}$, $p_{t0}$ and $\rho_0$ are the initial radial pressure, tangential pressure and energy density in static equilibrium
satisfying TOV. The first and the last term inside the square bracket respectively the anisotropic and relativistic correction respectively.
These are being positive quantities, increases the unstable range of the adiabatic index~\cite{Herrera1979,Chan1993}. Using the above relation we shall get the following expressions from our model as
\begin{eqnarray}\label{5.4.2}
&\qquad\hspace{-7cm} \Gamma _{{r}}=\frac{\rho+{p_r}}{p_r}\frac{dp_r}{d\rho}, \nonumber \\
&\qquad\hspace{-0.8cm} ={\frac {96 \left( n-1 \right) B_{{g}} \left( n+ \frac{2}{3}
 \right) \pi {r}^{6n+6}+ \left( -12\,{n}^{2}-12n \right) {r}^{6
n+4}-4\rho_{{1}}}{288 \left( n+\frac{2}{3} \right) B_{{g}}\pi \left(
n+ \frac{1}{2} \right) {r}^{6n+6}+ \left( -9{n}^{2}-9\,n \right) {r}^{6n+
4}-3\rho_{{1}}}}, \nonumber \\ \\ \label{5.4.3}
&\qquad\hspace{-7cm} \Gamma _{{t}}=\frac{\rho+{p_t}}{p_t}\frac{dp_t}{d\rho}\nonumber \\
&\qquad\hspace{-0.2cm} ={\frac {2 \left( \Gamma _{{1}}-\frac{1}{2}v_{{3}}{R}^{6n
+4}+{n}^{2}{r}^{6n+4} \right)  \left[ \rho_{{1}} \left( n-1 \right)
{r}^{-\left(4+6n\right)}+\Gamma _{{2}}-64B_{{g}}\pi {r}^{2}-p_{{1}} \right]
}{3 \left( \Gamma _{{3}}+{r}^{6n+4}n \right)  \left( \rho_{{1}}
 \left( n+2 \right) {r}^{-\left(4+6n\right)}+\Gamma _{{4}} \right) }},  \nonumber \\
\end{eqnarray}
where \\
$\Gamma _{{1}}=\frac{1}{2} \left( v_{{1}}-v_{{2}} \right)  \left( n+2 \right) $,\\
 $\Gamma _{{2}}=-9{n}^{2}+ \left( -64B_{{g}}\pi {r}^{2}-9 \right)n$,\\ $\Gamma _{{3}}= \left( 18{n}^{3}+27{n}^{2}+9n \right) {R}^{6n+4}-v_{{1}}+v_{{2}}$,\\ $\Gamma _{{4}}=80{r}^{2} \left( n+2/5 \right) B_{{g}}\pi -p_{{1}}$.

\begin{figure}[!htp]
\centering
    \includegraphics[width=6cm]{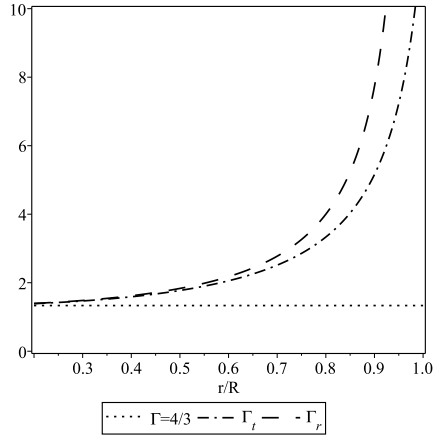}
 \caption{Variation of adiabatic index with the fractional radial coordinate $r/R$ for $LMC\,X-4$}\label{adia}
\end{figure}

To see the behaviour of the adiabatic index we have plotted $\Gamma_r$, $\Gamma_t$ with the fractional radial coordinate $\frac{r}{R}$ in Fig.~\ref{adia}. The figure shows that both $\Gamma_r,\Gamma_t > \frac{4}{3}$ everywhere within the stellar interior. This graphical representation indicates the stable model.

\section{Mass-radius relation and redshift}\label{sec6}
Buchdahl~\cite{Buchdahl1959} derived an upper limit for maximum allowed mass to radius ratio for a static spherically symmetric perfect fluid star, as $\frac{2M}{R}<\frac{8}{9}$. Later on Mak et al.~\cite{Mak2001} generalized it for a charged sphere. The mass of the star is given by
\begin{eqnarray}\label{6.1}
& \qquad m \left( r \right) =\frac{9r \left[ c_{{2}}{r}^{-\left(4+6n\right)}+{n}^{2}+ \left( \frac{16}{3}B_{{g}}\pi {r
}^{2}+1 \right) n+{\frac {32}{9}}B_{{g}}\pi {r}^{2} \right]}{18{n}^{2}+30n+12},\nonumber \\
\end{eqnarray}
where $c_{2}= \left( n+1 \right)  \left( n+ \frac{2}{3}\right) c_{1}$. This turns out to be $c_{2}= (7/4) c_{1}$ for our prescription of $n=1/2$. From Eq.~(\ref{4.6}) we immediately get the radius of the present compact star as $R = 4M$. This can be written in the form $2M/R=0.5$ which is well bellow the Buchdahl condition $2M/R<0.88$~\cite{Buchdahl1959} and thus provides the stability of the present model in terms of the mass-radius ratio.

Let us now define the compactification factor of the compact stellar system as
\begin{eqnarray}\label{6.2}
& \qquad u\left(r\right)=\frac{m}{r} =\frac{9 \left[ c_{{2}}{r}^{-\left(4+6n\right)}+{n}^{2}+ \left( \frac{16}{3}B_{{g}}\pi {r
}^{2}+1 \right) n+{\frac {32}{9}}B_{{g}}\pi {r}^{2} \right]}{18{n}^{2}+30n+12}.\nonumber \\
\end{eqnarray}

Therefore the gravitational redshift is defined as
\begin{equation}\label{6.3}
\qquad Z=(1-2u)^{-\frac{1}{2}}-1.
\end{equation}

\begin{figure}[!htp]
\centering
    \includegraphics[width=6cm]{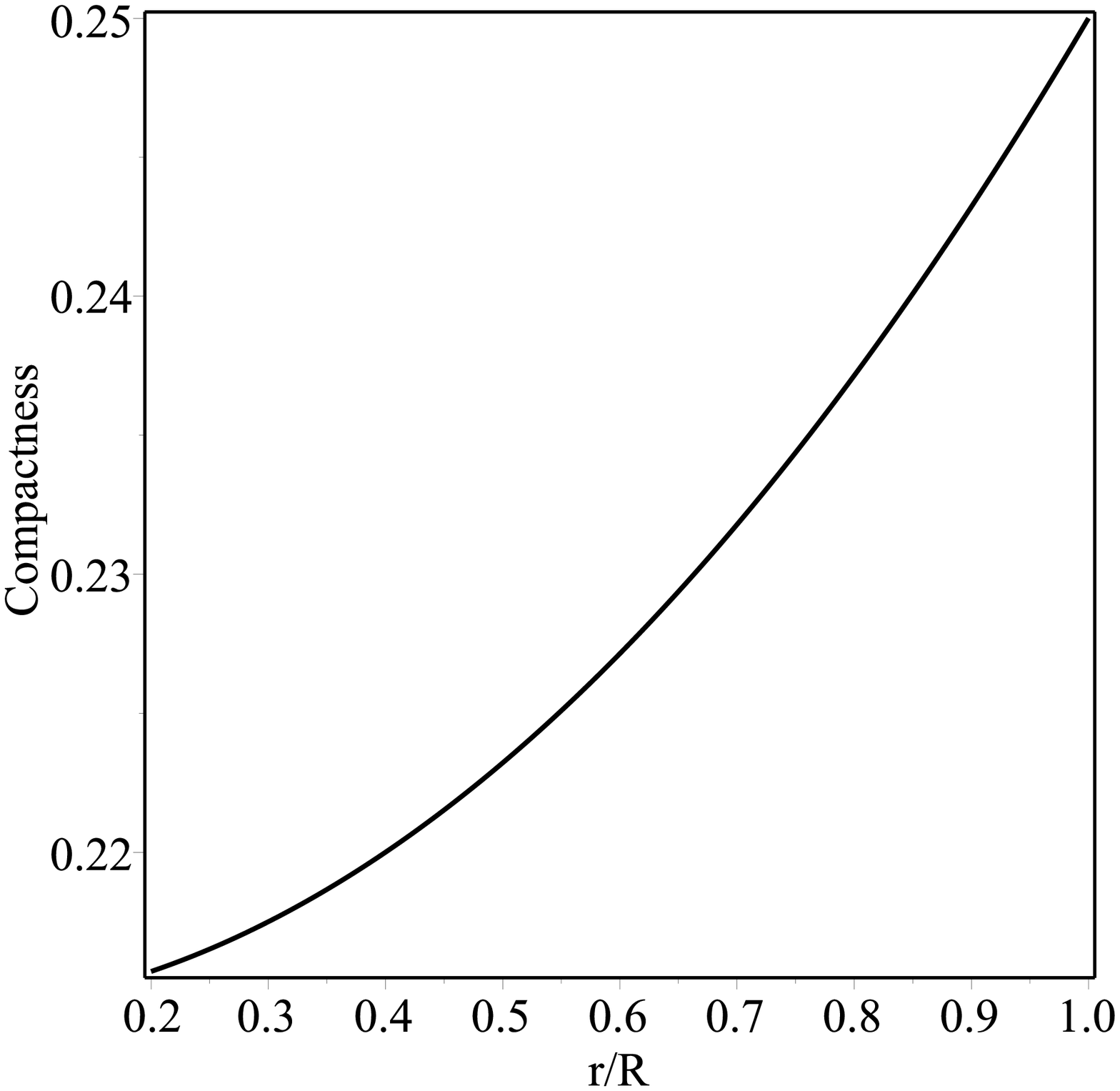}
    \includegraphics[width=6cm]{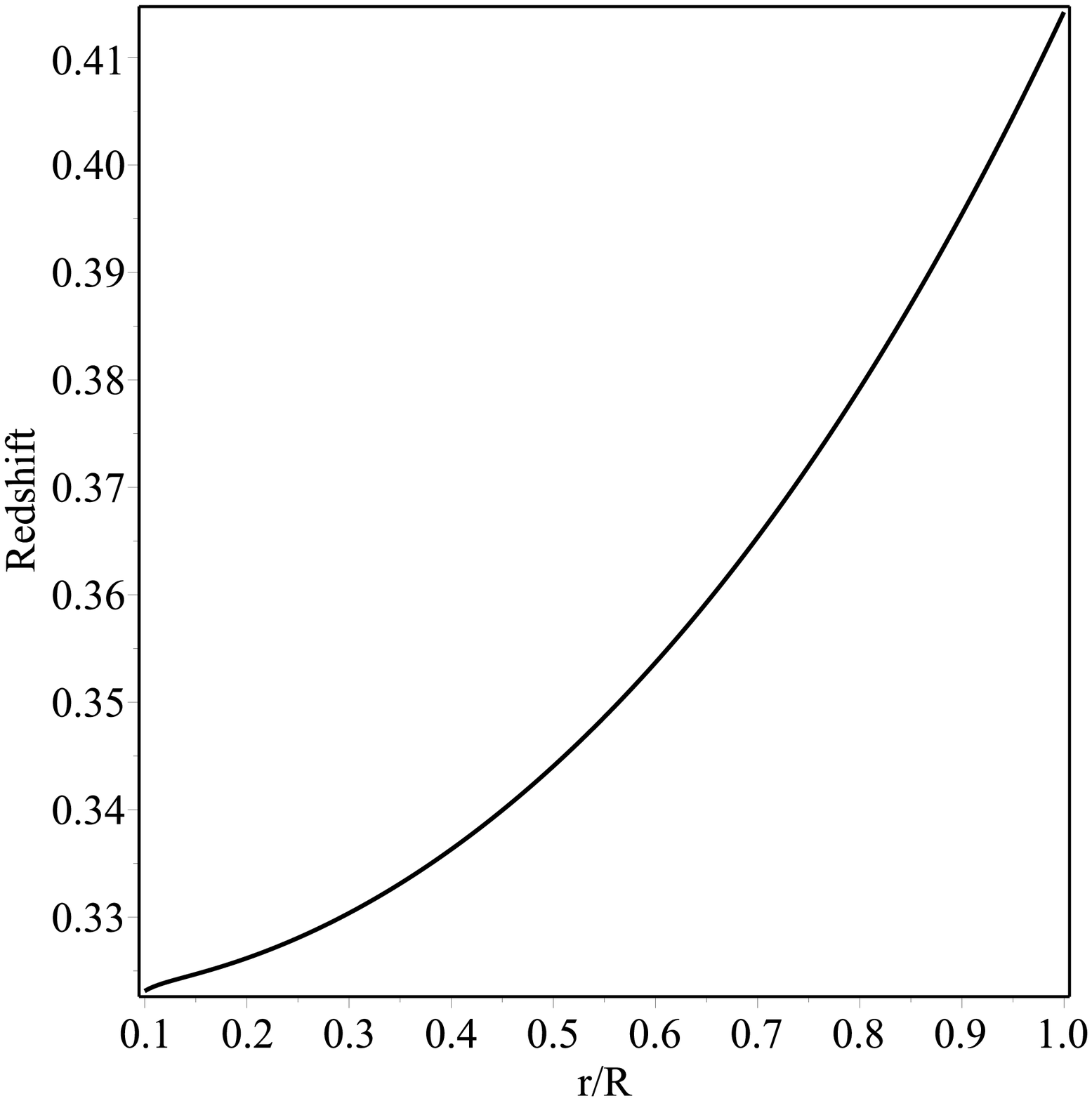}
               \caption{Variation of compactification factor (left panel) and redshift (right panel) with the fractional radial coordinate $r/R$ for $LMC\,X-4$} \label{redshift}
\end{figure}

Fig.~\ref{redshift} shows that the compactification factor increases non-linearly with the fractional radial coordinate and it satisfied Buchdahl limit. Also, the variation of redshift with respect to the fractional radial coordinate is featured in Fig.~\ref{redshift}.

\section{Discussion and conclusion}\label{sec7}
In this paper we have studied different features of a strange star through Tolman $V$ metric potential. With this potential and MIT bag EOS we have presented many interesting physical features of the strange stars and featured variation of the different physical parameters with the radial coordinate graphically. 

However, the most important result of the present model is that with the help of this model we can predict exact values of radius of the different strange stars and bag constant, which are featured in Table~\ref{Table1}. We find from this table that the surface density of different stars are much higher than the normal nuclear density $2.3\times{{10}^{14}}$~$gm/{cm}^3$ and such high density confirms the presence of quark matter inside the discussed stars. Also, we have derived a wide range of values of the bag constant. It have been shown by some authors~\cite{Farhi1984,Alcock1986} that the bag constant should be $55-75 Mev/(fm)^3$. But recent experimental results of CERN-SPS and RHIC confirms the possibility of a wide range of values of the bag constant. This is possible for a density dependent bag model~\cite{Burgio2002}. From the table it is very clear that the value of the bag constant increases with the increasing values of the density of the stellar systems. A density dependent bag constant have employed by Bordbar et al.~\cite{Bordbar2012} to model magnetized strange quark stars and obtained similar results as in the present model.

\begin{table}[!htp]
	\centering
\caption{Predicted Physical parameters where $1~{{M}_{\odot}}$=1.475~km for $G=c=1$. The complete data set is valid for $n=\frac{1}{2}$}\resizebox{\columnwidth}{!}{
\begin{tabular}{cccccccccccc}
 \hline Strange & Observed  & Predicted & \hspace{0.5cm}${\rho}_{s}$  & \hspace{0.3cm}${{B_g}}$  \\
Stars  & Mass $({{M}_{\odot}})$ & Radius (km) & (${gm/cm}^3$) & \hspace{-0.2cm}($MeV/{fm}^3$) \\
\hline $PSR~J~1614-2230$ & $1.97 \pm 0.04$ \cite{Demorest2010} & $11.623 \pm 0.236$  & $2.546 \times {10}^{14} $ & $ 35.704$\\
\hline $Vela~X-1$  & $1.77 \pm 0.08$\cite{Dey2013} &  $10.443 \pm 0.472$  & $ 3.152 \times 10^{14} $ & $44.204$\\
\hline $PSR~J~1903+327$ & $1.667 \pm 0.021$\cite{Dey2013} &  $ 9.835 \pm 0.120$ & $3.556 \times 10^{14}$ & $49.872$  \\
\hline $Cen~X-3$ & $1.49 \pm 0.08$\cite{Dey2013} &  $8.791 \pm 0.472$ & $4.459 \times 10^{14}$ & $62.528$  \\
\hline $LMC~X-4$ & $1.29 \pm 0.05$\cite{Dey2013} &  $7.611 \pm 0.292$ & $5.954 \times 10^{14}$ & $83.497$  \\
\hline $4U~1538-52$ & $0.87 \pm 0.07$\cite{Dey2013} &  $5.133 \pm 0.412$ & $13.08 \times 10^{14}$ & $183.429$  \\
\hline $SMC~X-1$ & $1.04 \pm 0.09$ \cite{Dey2013} &  $6.136 \pm 0.532$ & $9.146 \times 10^{14}$ & $128.268$  \\
\hline $Her~X-1$ & $0.85 \pm 0.15$\cite{Dey2013} &  $5.015 \pm 0.884$ & $13.70 \times 10^{14}$ & $192.119$  \\
\hline $4U~1820-30$ & $1.58 \pm 0.06$\cite{Guver2010b} &  $9.322 \pm 0.356$ & $3.960 \times 10^{14}$ & $55.539$\\
\hline $4U~1608-52$ & $1.74 \pm 0.14$\cite{Guver2010a} &  $10.266 \pm 0.828$ & $3.260 \times 10^{14}$ & $45.716$\\
\hline \label{Table1}
\end{tabular}}
\end{table}

Some other salient features from our study can be discussed as follows:

(i) We have studied our model for different values of $n$ as $n=1/2$, $n=3/5$, $n=4/5$ and $1$ but we find  physically valid solution for $n=1/2$ only which is the same as assumed by Tolman~\cite{Tolman1939}. 

(ii) We know that for a physically acceptable stellar model the metric potentials should be free from singularities inside the stellar structure. Our model satisfies this condition as at the centre metric potentials are giving finite values as ${e^{\nu}|_{r=0}}=0$  and ${e^{\lambda}}|_{r=0}>1$.

(iii) Though the parameters, like density and pressures, are singular at the origin, but if we take the ratio of the density with radial and tangential pressure separately, the singularity does not appear at the limit $r \rightarrow 0$. In this case we can have a linear relation between the pressures  with the density  at the centre as $ p_r=\rho$. Moreover, this is not geometrical singularity as the metric potentials are finite at the center. So the main reason of appearing the singularity is that the EOS is inadequate at the center. The highly ultra dense fluid at the center is not even compatible with the MIT bag EOS. Hence this issue can be resolved considering a core at the center of the stellar system where fluid distribution follows different EOS, for example $p_r=\rho$, a stiff fluid EOS as discussed earlier. The variation of the EOS parameters  are shown in Fig~\ref{eos}.

(iv) Through out the stellar distribution anisotropic force is positive (i.e., $p_t>p_r$) which helps to construct a more massive stellar structure.

(v) Our model satisfies all the energy conditions. The graphical representation of TOV equation shows that the stellar structure is in equilibrium under gravitational, anisotropic and hydrostatic forces. For our model causality condition and Herrera's cracking condition holds simultaneously representing a stable configuration. Also variation of adiabatic index is in expected region.

(vi) In the present study we have opted for the value of $n=1/2$ as other values do not yield physically viable results. Therefore, from Eq.~(\ref{4.6}) we get the radius of the present compact star as $R = 4M$. This can be written in the form $2M/R=0.5$ which fulfils the Buchdahl condition $2M/R<0.88$~\cite{Buchdahl1959} in connection to the mass-radius relationship. For the ultradense strange star $LMC~X-4$ we find the value of compactness as $0.25$ which also satisfies the Buchdahl limit. The surface redshift in this case is $0.41$.

Overall, by using Tolman $V$ metric potential we have represented a stable anisotropic and compact stellar configuration which satisfies all the physical conditions and gives exact numerical values of some of the physical parameters.

\section*{Acknowledgments}
SR is thankful to the authority of the Inter-University Centre for Astronomy and Astrophysics, Pune, India and Institute of
Mathematical Sciences, Chennai, India for providing Visiting Associateship under which a part of this work was carried out. 
DS is also grateful to the authority of IUCAA for the hospitality during the visit.

\end{document}